\documentstyle[editedvolume,numreferences]{crckapb}

\begin{opening}
\title{SPECIAL K\"AHLER GEOMETRY}
\subtitle{Does there Exist a Prepotential?}
\author{BEN CRAPS}
\author{FREDERIK ROOSE}
\author{WALTER TROOST}
\author{ANTOINE VAN PROEYEN}
\institute{Instituut voor Theoretische Fysica\\Katholieke Universiteit Leuven, 
  B-3001 Leuven, Belgium}
\end{opening}

\newcommand{\ft}[2]{{\textstyle\frac{#1}{#2}}}
\newsavebox{\uuunit}
\sbox{\uuunit}
    {\setlength{\unitlength}{0.825em}
     \begin{picture}(0.6,0.7)
        \thinlines
        \put(0,0){\line(1,0){0.5}}
        \put(0.15,0){\line(0,1){0.7}}
        \put(0.35,0){\line(0,1){0.8}}
       \multiput(0.3,0.8)(-0.04,-0.02){12}{\rule{0.5pt}{0.5pt}}
     \end {picture}}
\newcommand {\unity}{\mathord{\!\usebox{\uuunit}}}
\newcommand{\Ka}{K\"ahler}

\newcommand  {\Rbar} {{\mbox{\rm$\mbox{I}\!\mbox{R}$}}}

\newsavebox{\zzzbar}
\sbox{\zzzbar}
  {\setlength{\unitlength}{0.9em}
  \begin{picture}(0.6,0.7)
  \thinlines
  \put(0,0){\line(1,0){0.6}}
  \put(0,0.75){\line(1,0){0.575}}
  \multiput(0,0)(0.0125,0.025){30}{\rule{0.3pt}{0.3pt}}
  \multiput(0.2,0)(0.0125,0.025){30}{\rule{0.3pt}{0.3pt}}
  \put(0,0.75){\line(0,-1){0.15}}
  \put(0.015,0.75){\line(0,-1){0.1}}
  \put(0.03,0.75){\line(0,-1){0.075}}
  \put(0.045,0.75){\line(0,-1){0.05}}
  \put(0.05,0.75){\line(0,-1){0.025}}
  \put(0.6,0){\line(0,1){0.15}}
  \put(0.585,0){\line(0,1){0.1}}
  \put(0.57,0){\line(0,1){0.075}}
  \put(0.555,0){\line(0,1){0.05}}
  \put(0.55,0){\line(0,1){0.025}}
  \end{picture}}

\def\IP{\relax{\rm I\kern-.18em P}}
\def\Im{{\rm Im ~}}
\def\Re{{\rm Re ~}}
\newcommand{\beq}{\begin{equation}}
\newcommand{\eeq}{\end{equation}}
\newcommand{\IM}{{\rm Im~}}

\newcommand{\N}{\mbox{$\cal N$}}
\newcommand{\dee}[1]{\mbox{$\partial_{#1}$}}
\newcommand{\Sp}[1]{\mbox{$Sp\left( #1,\Rbar \right) $}}

\newcommand{\sinprod}[2]{\mbox{$\langle #1 , #2 \rangle$}}
\newcommand{\eqn}[1]{(\ref{#1})}

\newcommand{\Dee}[1]{\mbox{${\cal{D}}_{#1}$}}

\newcommand{\ZK}{\mbox{$Z^{K}$}}
\newcommand{\ZL}{\mbox{$Z^{L}$}}

\newcommand{\mtrx}[4]{\left(
  \begin{array}{cc}
    {#1} & {#2} \\
    {#3} & {#4}
  \end{array}\right)}

\begin{document}

\begin{abstract}
A symplectically invariant definition of special K\"ahler geometry,
proposed in \cite{whatskg}, is discussed. Certain aspects hereof are 
illustrated by means of Calabi-Yau moduli spaces. 
\end{abstract}

\section{Introduction}

Special (K\"ahler) geometry \cite{DWVP} is, by definition, the geometry of 
vectormultiplet
scalars in $N=2$ supergravity. However, one would like to define this geometry
only referring to these scalars, not to any other fields.
Therefore one needs to know the most
general way of coupling vectormultiplets to supergravity. Originally,
supergravity actions for vectormultiplets were constructed using a holomorphic
`prepotential'. As turned out later,
duality transformations can lead to actions for which a
prepotential does not exist \cite{f0art}. 
In \cite{whatskg} a formulation (`definition')
of special
geometry was given which is manifestly invariant under duality transformations.
It was proved that this formulation is equivalent to the original
one, in the sense
 that it is always possible to perform a duality transformation such
that a prepotential exists in the dual formulation of the theory. Moreover, it
describes all presently known examples of special geometry.      
All constraints imposed in this definition have a nice physical interpretation
(related to duality invariance and positivity of the kinetic energy), except
for one constraint in the special case of only one vectormultiplet. This
exception suggests that one could try to construct a more general supergravity
theory for one vectormultiplet, one that could not be encoded in a holomorphic
prepotential.

The aim of this contribution is to review this evolution, simplifying the 
discussion by omitting a lot of
details and concentrating on the points essential for the interpretation of the
constraints imposed in \cite{whatskg}.

\section{$N=2$ supergravity coupled to vectormultiplets}

The (abelian) theories we are going to consider describe a SUGRA multiplet 
(whose bosonic
dynamical degrees of freedom are the graviton and a vector, the graviphoton) and
$n$ vectormultiplets (each a vector and a complex scalar). Thus we have $n+1$
vectors (described by ${\cal F}_{\mu\nu}^I$, $I=0,\ldots,n$) and $n$ complex
scalars $z^\alpha$, $\alpha=1,\ldots,n$.

The scalars are interpreted as coordinates on a manifold ${\cal M}$. This
manifold allows projective coordinates 
\[Z^I(z^\alpha) \sim e^{f(z^\alpha)}\, Z^I(z^\alpha)\ .\]
There exists a holomorphic function $F(Z^I)$, homogeneous of second degree,
named the `prepotential', from which the following $2n+2$ component vector 
can be defined:
\begin{equation} \label{v}
  v\equiv
  \left( \begin{array}{c} Z^I \\ \partial F/\partial Z^I \end{array}\right)\equiv
  \left( \begin{array}{c} Z^I \\ F_I \end{array}\right)\ .
\end{equation}

The kinetic term ${\cal L}_0$ of the scalars is given by
\begin{equation}
  {\cal L}_0=-e\,g_{\alpha\bar\beta}\,\partial_\mu z^\alpha\,\partial^\mu \bar
  z^\beta \ ,
\end{equation}
where $e$ is the determinant of the vierbein and $g_{\alpha\bar\beta}$ is a
K\"ahler metric with K\"ahler potential
\begin{equation}  \label{K}
K=-\ln[i<\bar v,v>]\ .
\end{equation}
Here the bracket denotes the symplectic inproduct
\[
  \langle V , W \rangle \equiv V^T \Omega W 
  \qquad\mbox{with}\qquad
  \Omega=\pmatrix{0&\unity  \cr -\unity  &0\cr}\ .
\]
The metric $g_{\alpha\bar\beta}$, describing the geometry of the scalar
manifold ${\cal M}$, is called {\it special K\"ahler}. 

The vector kinetic term is
\begin{equation}
  {\cal L}_1=e\left(
{\ft14} (\Im {\cal N}_{IJ}){\cal F}_{\mu\nu}^I
{\cal F}^{\mu\nu J}
-{\ft i8} (\Re {\cal N}_{IJ})
\epsilon^{\mu\nu\rho\sigma}{\cal F}_{\mu\nu}^I
{\cal F}_{\rho\sigma}^J\right)\ ,   \label{genL01}
\end{equation}
where
\[
\N_{IJ}(Z)= \bar{F}_{IJ}(\bar Z)+2i\, 
\frac{\IM F_{IK}(Z)\: \IM F_{JL}(Z)\: \ZK\,\ZL}
          {\IM F_{KL}(Z)\:\ZK\ZL} \ .
\]

The combination of equations of motion and Bianchi identities
  is invariant under symplectic {\it duality transformations}
\begin{eqnarray}
  v&\rightarrow &\mtrx{A}{B}{C}{D} \, v\\
 {\cal N}&\rightarrow&(C + D{\cal N})(A+B{\cal N})^{-1}\ ,
\end{eqnarray}          
with $\mtrx{A}{B}{C}{D} \in \Sp{2n+2}$.
These transformations leave $g_{\alpha\bar\beta}$ invariant.

As an {\it example}, consider $n=1$ and $F(Z^0,Z^1)=-iZ^0Z^1$, with $Z^0=1$ and
$Z^1=z$. Thus
\[
  v=\left( \begin{array}{c} Z^I \\ F_I \end{array}\right)=
    \left( \begin{array}{c} 1\\ z\\ -iz\\-i\end{array}\right)\ ,
\]
$e^{-K}=2(z+\bar z)$ and $g_{z\bar z}=(z+\bar z)^{-2}$. Now consider the
following symplectic transformation:
\[
  v\rightarrow\tilde v=\left(
  \begin{array}{cccc} 1&0&0&0\\0&0&0&-1\\0&0&1&0\\0&1&0&0\end{array} \right)v
  =\left( \begin{array}{c} 1\\i\\-iz\\z\end{array}\right)
\]
It is clear that no prepotential exists for $\tilde v$, since $\tilde v$ is not
of the form \eqn{v}
(its lower part is independent of its upper one).

\section{Symplectically invariant definition}
\label{ss:symp}

We define a special K\"ahler metric as a K\"ahler metric\footnote{We include
positivity in the definition of a K\"ahler metric.} for which there exists
a $2n+2$ component vector $v(z^\alpha)$ such that \eqn{K} is satisfied and 
\begin{equation} \label{vdv}
  <v,\Dee{\alpha}v>=0 \ ,
\end{equation}
where $\Dee{\alpha}\equiv\dee{\alpha}+(\dee{\alpha}K)$ is covariant under
$v\rightarrow e^{f(z)}\,v$. 

Using the notation $v=\left( \begin{array}{c} Z^I \\ F_I \end{array}\right)$,
where $F_I$ need not be the derivative of a prepotential, we can write down the
following manifestly symplectically covariant form of \N:
\begin{equation} \label{N}
  \N_{IJ}=(\begin{array}{cc}\Dee{\bar\alpha}\bar F_I&F_I\end{array})
  (\begin{array}{cc}\Dee{\bar\alpha}\bar Z^J&Z^J\end{array})^{-1}\ .
\end{equation}

It was proved in \cite{whatskg} that this definition is equivalent to the
prepotential--definition and that it includes all (presently) known
supergravity actions.

Let us remark that the matrix $\left(\begin{array}{cc}\Dee{\bar\alpha}\bar Z^J&
Z^J\end{array}\right)$ is always invertible, whereas the invertibility of
$\left(\begin{array}{cc}\Dee{\alpha} Z^J& Z^J\end{array}\right)$ is equivalent
to the existence of a prepotential for $v$.

\section{Interpretation of the constraint $<v,\Dee{\alpha}v>=0$}
\label{ss:interpr}

The constraint \eqn{vdv} implies the positivity of the following matrix:
\begin{equation}  \label{posmat}
\pmatrix{\Dee{\alpha}Z^I\cr \bar Z^I}i\,e^K({\cal N}- {\cal N}^{\dagger})_{IJ}
\pmatrix{\Dee{\bar \beta}\bar Z^J& Z^J}=
  \left(\begin{array}{cc}g_{\alpha\bar\beta}&i<\Dee{\alpha}v,v>\\
  i<\bar v,\Dee{\bar\beta}\bar v>&e^{-K}\end{array}\right),
\end{equation}
which in turn implies both that $\left(\begin{array}{cc}\Dee{\bar\alpha}
\bar Z^J& Z^J\end{array}\right)$ is invertible 
(guaranteeing that \N\ is well-defined) and, using the symmetry of \N,
that $\Im\N$ is negative (ensuring the positivity of the vector kinetic
energy).

On the other hand, the symmetry of \N\ defined in \eqn{N} (needed for duality
invariance) is equivalent to 
\begin{equation}  \label{dvdv}
  <\Dee{\alpha}v,\Dee{\beta}v>=0\ .   
\end{equation}
Whereas \eqn{vdv} implies \eqn{dvdv}, the reverse only holds for $n>1$.

This raises the question whether the constraint \eqn{vdv} can be relaxed for
$n=1$ to the combination of the positivity of \eqn{posmat}, and \eqn{dvdv}. 
Solving this question
means verifying whether it is possible to construct an $N=2$ supergravity 
theory for a single vectormultiplet with a vector $v$ satisfying both
the positivity of
\eqn{posmat} and \eqn{dvdv} but not \eqn{vdv}. This problem has not been solved
yet.  

\section{Special geometry and Calabi-Yau moduli spaces}

The low-energy description of a type II string theory compactified on
a Calabi-Yau threefold is a $d=4, N=2$ supergravity theory. As such
the (vector multiplet) scalar sector involved in such compactifications is expected to
yield a special \Ka\  manifold. In this section we will outline how
the various constraints of special \Ka\  geometry are realised in this
concrete setting.

The interpretation of the special geometry relations goes basically via the
following isomorphism
\begin{equation} \label{cohsym} \begin{array}{ccccc}
H^3(X) & \rightarrow && \Sigma & \\
\omega & \mapsto & v_\omega &=& 
\left(\begin{array}{c}\int_{A^I} \omega \\ \int_{B_I} \omega \end{array}
 \right)\ ,
\end{array} 
\end{equation}
where $X$ is some generic CY threefold and $\{A^I,B_I\}$ a canonical basis
of 3-cycles. $\Sigma $ is a $2 h^{2,1} + 2$ complex dimensional vector
space. The threeform cohomology may be endowed with the
sesquilinear form $Q$:
\[ Q(\eta, \omega) = i\int_X \eta \wedge \bar\omega\ ,
\]
which corresponds to a symplectic form $i\sinprod{v_\eta}{v_\omega}$
on $\Sigma$. Note that a symplectic transformation on $\Sigma$
correspond to a change of canonical homology basis. The form $Q$ can be
shown \cite{grifharr} to enjoy the
following properties:
\begin{enumerate}
\item It is block-diagonal in an appropriate cohomology basis. More
explicitly, when expressed with respect to any cohomology basis adapted
to the Hodge decomposition $H^3 = H^{3,0}\oplus  H^{2,1}\oplus
H^{1,2}\oplus H^{0,3}$, $Q$ takes the following form
\[ \left( 
\begin{array}{cccc}
H & 0 & 0 & 0 \\
0 & -H^*_{\bar\alpha\beta} & 0 & 0 \\
0 & 0 & H_{\alpha\bar\beta} & 0 \\
0 & 0 & 0 & -H 
\end{array} \right)\ . \]
\item Q is positive-definite on $H^{3,0}\oplus H^{1,2}$.
\end{enumerate}

Now consider a family of CY threefolds with varying complex structure
and fixed \Ka\  class. In the generic case $h^{2,1}$
complex parameters $z^\alpha$ specify one particular member in this
family. Furthermore fix a complex structure on this space of moduli such
that the unique holomorphic threeform $\Omega$ on $X$ depends only
holomorphically on these moduli $z^\alpha$. As small variations of a $(p,q)$ form
give at most $(p\pm 1, q \mp 1)$ forms, we find
\[
\begin{array}{ccccc}
\partial_\alpha \Omega &=& -\kappa_\alpha \Omega &\oplus & \Omega_\alpha
\ , \\
&\in & H^{3,0}&\oplus& H^{2,1}\ .
\end{array}
\]
Thus we may define $\Dee\alpha \Omega \equiv [\partial_\alpha +
\kappa_\alpha ] \Omega$. Moreover this set can be shown to constitute
a basis for $H^{2,1}$. We define further $\Dee{\bar\alpha} v \equiv
\partial_{\bar\alpha} v = 0$. 
Under the mapping (\ref{cohsym}) the threeforms
$\{\Omega, \Omega_\alpha\} $ are mapped to a corresponding set $\{v,
\Dee{\alpha} v\}$ of symplectic vectors in $\Sigma$, with the mentioned
properties induced.   

Putting things together we are now in a position to state the main
results of this whole setup. After specifying some arbitrary canonical
homology basis we define $v$ to be the image of
the unique holomorphic threeform $\Omega$  under the mapping (\ref{cohsym}).
This object $v$ is the one central object in section~\ref{ss:symp}. 
Furthermore $\Dee{}$ is defined as in the above paragraph.  

The various properties of $Q$ may now be properly translated into
relations satisfied by $v$ and $\Dee{\alpha} v$. 
The quantity $-i\sinprod{v}{\bar v}$ is strictly positive as is 
correspondingly the matrix entry $H$ in the bilinear form $Q$ on
cohomology. As a result we may define a real \Ka\ potential through
$-i\sinprod{v}{\bar v} = e^{-K}$. It is an easy exercise to check that
the \Ka\  potential thus defined appears in the derivative $\Dee{\alpha}
= \partial_\alpha + (\partial_\alpha K)$, precisely as in section
\ref{ss:symp}.
The positivity of $Q$ on $H^{1,2}$
is equivalent to the positivity of the metric which we define to be
$g_{\alpha\bar\beta} = -i\sinprod{\Dee{\alpha} v}{\Dee{\bar\beta}\bar
v}$. This metric coincides with the one obtained from the \Ka\ 
potential, $g_{\alpha\bar\beta} = \partial_\alpha\partial_{\bar\beta}K$. The
derivative $\Dee{}$ is covariant under holomorphic rescalings $v
\rightarrow e^{f(z)}v$, which in turn correspond to 
\Ka\  transformations on the moduli space. 
From the block-diagonal form of $Q$ in the specified
basis the following relations
are immediate consequences:
\begin{eqnarray}
\sinprod{v}{\Dee{\alpha} v} &=&  0\ , \\
\sinprod{v}{\Dee{\bar\alpha}\bar v} &=& 0 \ .
\end{eqnarray}

As to the invertibility of the matrix $\left(\begin{array}{cc}\Dee{\bar\alpha}\bar Z^J&
Z^J\end{array}\right)$ in section \ref{ss:symp}, first
notice that it corresponds to 
$\left(\begin{array}{cc}\int_{A^J}\bar\Omega_{\bar\alpha}& \int_{A^J} \Omega
\end{array}\right)$ in the present context. The positivity of $Q$ on $H^{3,0}\oplus H^{1,2}$ guarantees its
invertibility \cite{whatskg}. Replacing $H^{1,2}$ with $H^{2,1}$, no similar
statement can be made, as a result of which $\left(\begin{array}{cc}\Dee{\alpha} Z^J&
Z^J\end{array}\right)$, or equivalently
$\left(\begin{array}{cc}\int_{A^J}\Omega_{\alpha} & \int_{A^J} \Omega
\end{array}\right)$, 
may not be invertible. Via an appropriate symplectic rotation on
$\Sigma$ it is always possible to make it invertible,
though \cite{whatskg}.
A prepotential formulation of the special geometry of the CY moduli
space is absent if the above matrix is non-invertible. This non-invertibility
is due to a particular choice of 3-cycles.

Concerning section \ref{ss:interpr} it is obvious that in the present case the
stronger constraint $\sinprod{v}{\Dee{\alpha} v}$ is always obeyed, as
is implied by $Q$ and its properties once more. 

\acknowledgements
B.C. and F.R. would like to thank the organizers for the opportunity
to give these talks.
B.C. (Aspirant FWO), W.T. (Onderzoeksleider FWO) and A.V.P.
(Onderzoeksdirecteur FWO) thank the Fund for Scientific
Research--Flanders for financial support.
This work was supported by
the European Commission TMR programme ERBFMRX-CT96-0045.

\end{document}